\documentclass[twocolumn,showpacs,aps,floatfix]{revtex4}

\usepackage{graphicx}

\newcommand{\km}{\,\mathrm{km}}
\newcommand{\s}{\,\mathrm{s}}
\newcommand{\yr}{\,\mathrm{yr}}
\newcommand{\eV}{\,\mathrm{eV}}
\newcommand{\erg}{\,\mathrm{erg}}
\newcommand{\Mpc}{\,\mathrm{Mpc}}
\newcommand{\ud}{\mathrm{d}}

\begin{document}

\title{On the shape of the UHE cosmic ray spectrum
}
\author {Daniel De Marco \& Todor~Stanev}
\affiliation{
Bartol Research Institute, 
University of Delaware, Newark, DE 19716, USA
}
\widetext
\begin{abstract}
 We fit the ultra high energy cosmic ray spectra above $10^{19}\eV$
 with different injection spectra at cosmic ray sources that are
 uniformly and homogeneously distributed in the Universe. We conclude
 that the current UHE spectra are consistent with power laws of
 index $\alpha$ between 2.4 and 2.7. There is a slow dependence of
 these indices on the cosmological evolution of the cosmic ray
 sources, which in this model determines the end of the 
 galactic cosmic rays spectrum.
\end{abstract}

\date{\today}
\maketitle

\section{Introduction}
 The current results on the energy spectrum of the highest energy
 cosmic rays are not fully consistent. The two high statistics
 experiments, AGASA~\cite{AGASA} and HiRes~\cite{HiRes_PRL}, do
 not agree on the normalization of the ultra high energy cosmic ray
 (UHECR) spectrum. In addition, HiRes results are consistent with a
 GZK~\cite{GZK} suppression, and AGASA claims a spectrum extended to
 higher energy. With the current statistics  the differences are not
 very significant -- the number of events above $10^{20}\eV$ differs
 by less than 3$\sigma$~\cite{BDMO}.
 The normalizations of the spectra are quite different, especially in
 the common $E^3 \times \ud N/\ud E$ presentation, but  a renormalization
 of the energy assignment by 15--20\%, which is within the  reported 
 systematic uncertainty of the energy assignment, of both data sets
 leads to a good agreement~\cite{BDMO,TSRiken}. The high energy
 extension and exact normalization of the UHECR spectrum are thus not
 well known, but after the renormalization both experiments show the
 same spectral shape between $10^{18.5}$ and $10^{20}\eV$.

 There have been recently several
 attempts~\cite{BGG1,WB,WW,BGG2,L04,AB04,HiRes_APP,SS05} to explain
 this shape with different injection spectra of extragalactic protons
 after propagation to the observer from isotropically and homogeneously
 distributed sources.
 The assumed injection spectra and to certain extent the cosmological
 evolution of the sources determine the shape of the extragalactic
 cosmic ray spectrum at Earth.
 A subtraction from the observed cosmic ray spectrum in these models
 determines also the end of the galactic cosmic ray spectrum.
 There are two types of solutions. Flat injection spectra,
 $\ud N/\ud E = A E^{-2}$, are suggested in Refs.~\cite{WB,WW}. In this 
 case the galactic cosmic rays spectrum extends above $10^{19}\eV$.
 The other popular solution is to use much steeper injection spectra with
 spectral indices $\alpha$ = 2.6--2.7. Such solutions set the end of
 the galactic cosmic ray spectrum at lower energy.
 The assumption that the extragalactic cosmic rays have a large fraction
 of nuclei heavier than Hydrogen~\cite{APKGO05} modifies the 
 propagation process and can potentially initiate a new class of
 models, in which the composition at injection introduces more
 model parameters. 
 
 Cosmic ray data is not yet good enough to prove that any of these
 models are correct. Fitting the spectra with several different
 parameters and the uncertain knowledge of the cosmic ray composition
 at $10^{19}\eV$ makes all models plausible.

 We describe an attempt to fit the renormalized AGASA and HiRes
 spectra with injection spectra and cosmological evolution 
 parameters covering practically the whole phase space and to use
 the quality of the fits as a measure of the plausibility of
 the models. This involves several assumptions that, although
 used in many previous publications, may not be correct.
 The main one is that extragalactic cosmic rays are protons
 and that the renormalized experimental spectra represent correctly
 the shape of the UHECR spectrum. One of the fit parameters is
 the required extragalactic cosmic ray luminosity at present
 time. The uncertainty of the luminosity depends on the 
 arbitrary renormalization of the experimental data, as well
 on the cross correlation with other parameters and assumptions.

 This paper is organized as follows: Section 2 describes the fitting 
 procedures that we use, Section 3 gives the main results of the 
 fits, and Section 4 discusses the results, compares to other fits
 and derives the main conclusions from this research.

\section{Fitting the UHECR spectrum}

 It was shown in Refs.~\cite{BDMO,TSRiken} that a renormalization of
 about 15\% of the energy assignment of the AGASA and HiRes events,
 would
 bring the two spectra in very good agreement in the energy region below
 $10^{20}\eV$ in both normalization and shape. In Ref.~\cite{BDMO} it
 was also shown that the statistics of events above $10^{20}\eV$ is too
 small to achieve a conclusive result about the end of the UHECR
 spectrum. In this paper we study how the best fit to the injection
 spectrum depends on the source parameters: injection spectrum,
 luminosity and luminosity evolution with redshift. To do so, supported
 by the above-mentioned findings, we fit these injection parameters to
 the energy-shifted spectra in the energy range $10^{19}$--$10^{20}\eV$.
 We chose as lower bound $10^{19}\eV$ because we expect at this energy
 the particles to be mostly extra-galactic and $10^{20}\eV$ as higher
 bound because of the sparseness of data above this threshold.
 
 To calculate the expected spectrum from a isotropic homogeneous
 distribution of proton sources we use the analytical approach presented
 in Ref.~\cite{bg,BGG1}. In this approach all proton energy losses are
 included (redshift losses, pair and pion production losses) and they
 are all treated as continuous. This is the correct treatment for
 redshift losses and it is well suited for pair production and for pion
 production at large propagation distances. At small propagation
 distance, however, the large inelasticity of the pion production
 process produces large fluctuations in the expected fluxes that cannot
 be reproduced in the continuous energy loss approximation. A Monte
 Carlo simulation is better suited in this case. In the present paper,
 since pion production affects mostly the highest part of the energy
 spectrum, and we are only interested in the spectra below $10^{20}\eV$,
 we can safely use the continuous energy loss approximation for all
 energy loss processes.

 \begin{figure}[htb]
 \vspace*{-15pt}
   \centering
   \includegraphics[width=70truemm]{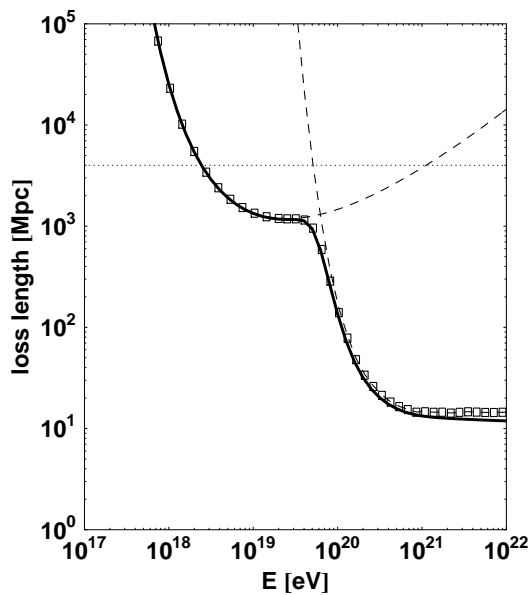}
 \vspace*{-15pt}
   \caption{Solid line: loss length for photo-pion and photo-pair
   production for protons as calculated in Ref.~\cite{BGG1}. The dashed
   lines report the separate contribution of the two processes. The
   dotted line shows the loss length for redshift losses. The squares
   represent the loss length of Ref.~\cite{Stanev:2000fb}.}\label{fig:ll} 
 \vspace*{-10pt} 
\end{figure}

 We assume the sources inject protons with a power-law spectrum,
 $E^{-\alpha}$, with a sharp cutoff at $E_\mathrm{max}=10^{21.5}\eV$.
 Changing the value of $E_\mathrm{max}$ does not appreciably affect the
 results in the energy region we are interested in. We assume the source
 emissivity to evolve as $L(z)=L_0(1+z)^m$, where $L_0$ is the present
 cosmic ray emissivity of the sources and $m=0$ corresponds to absence of
 evolution.
 There is no cutoff to the evolution of the luminosity because we are
 only interested in the region above $10^{18.5}\eV$ and in this region
 the contributions to the observed flux can come only from sources with
 $z\lesssim 0.5$.
 To calculate the energy losses we use the loss-lengths of
 Ref.~\cite{BGG1}, plotted in Fig.~\ref{fig:ll}, that were shown to be
 indistinguishable from the ones of Ref.~\cite{Stanev:2000fb} in the low
 energy region and within 15\% at high energy~\cite{BGG1}. We assume a
 $\Lambda$CDM universe with $\Omega_\Lambda=0.7$, $\Omega_m=0.3$ and
 $H_0=75\km\s^{-1}\Mpc^{-1}$.

 We explore the parameter region in $\alpha$ from 2.05 to 3.00 in steps
 of 0.05 and in $m$ from 0 to 4 in steps of 0.25. For each $(\alpha,m)$
 pair we calculate the expected flux and then the best fit
 emissivity, $L_0$, minimizing the $\chi^2$ indicator.
 To emulate the experimental energy resolution we include 30\% Gaussian 
 error distribution in the spectra after propagation.
 The effect of the energy resolution is to smooth the features
 produced by the propagation on the photon background and to lessen
 the GZK suppression~\cite{BDMO}.

 \begin{figure}[htb]
   \centering
   \includegraphics[width=85truemm]{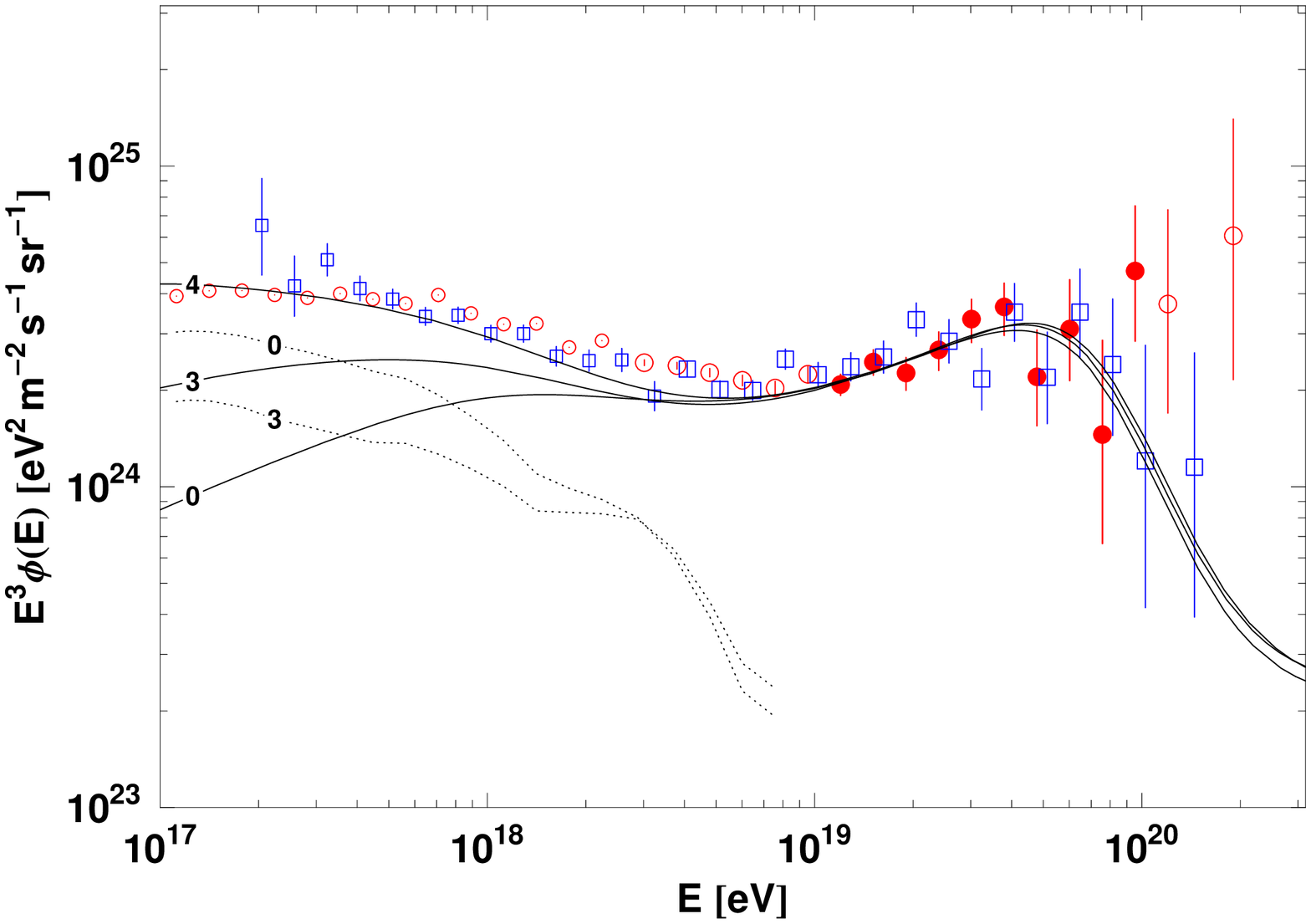}
   \includegraphics[width=85truemm]{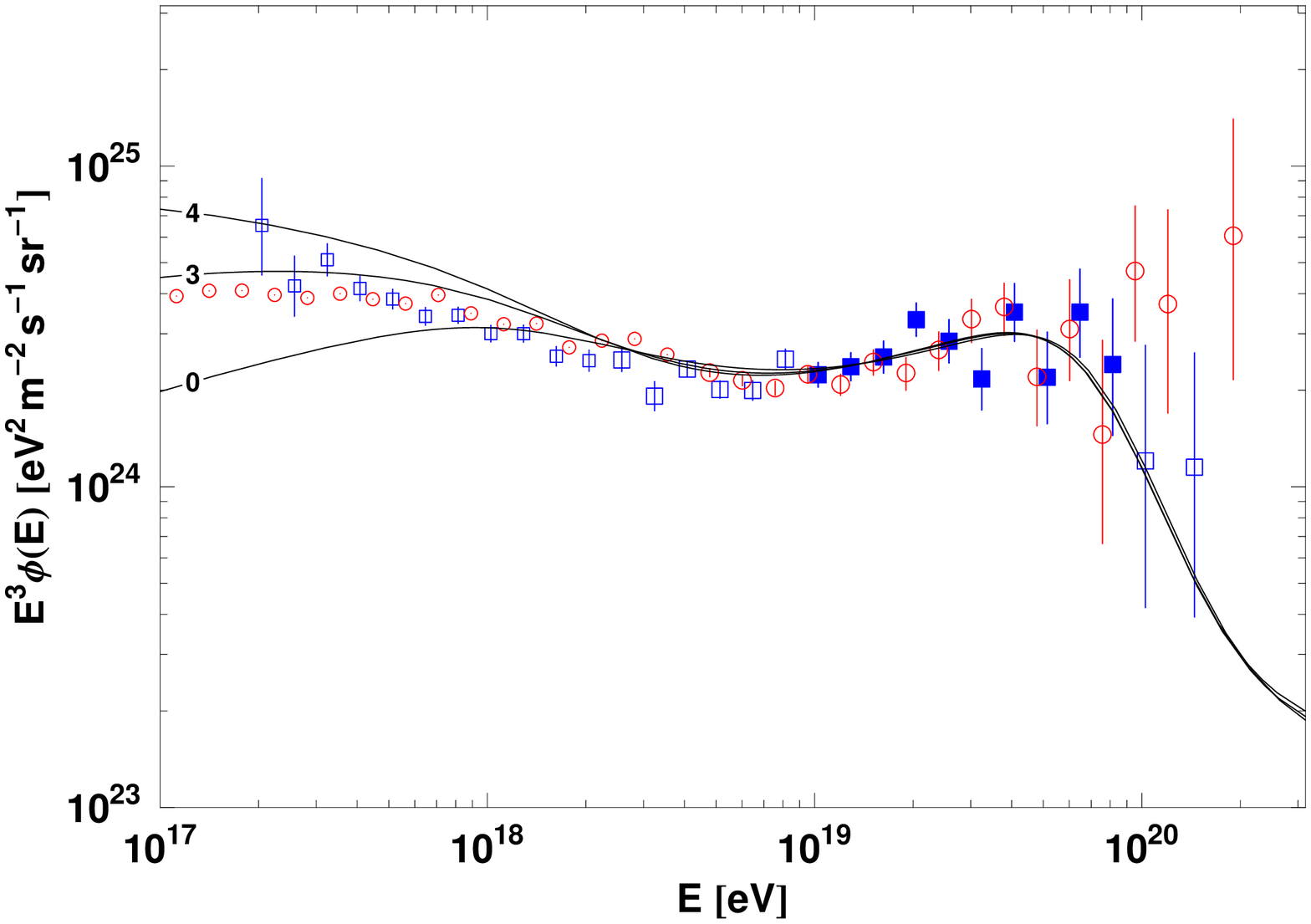}
   \caption{Upper panel: Best fits to the AGASA $-15\%$ dataset in the
   \protect$10^{19}$--$10^{20}\eV$ energy range. Squares with error
   bars: HiRes dataset. Circles with error bars: AGASA\&Akeno dataset.
   Filled symbols: points used in the fit. Smaller data points: data
   from HiRes2 and from Akeno (not shifted) to show the low-energy shape
   of the spectrum. Solid lines: best fits for different values of $m$.
   Dotted lines: galactic component needed in order to fit the spectrum
   at lower energy. The numbers attached to the lines indicate the value
   of $m$. Lower panel: best fit for the HiRes $+15\%$
   dataset.}\label{fig:spectra190} 
 \end{figure}

 For AGASA and Akeno we use the data from Ref.~\cite{AGASA}, whereas for
 HiRes 1\&2 we use the data from their website~\cite{HiResweb}
 which is very close to the published results~\cite{HiRes_PRL,HiRes_APP}.
 Following the suggestion of Ref.~\cite{BDMO} we shift the
 AGASA and HiRes energies respectively by $-15\%$ and $+15\%$, while we
 leave the Akeno energies unchanged~\cite{NaganoWatson}. To do the shift
 we proceed in the
 following way: we calculate $\frac{\ud N}{\ud \log_{10}(E)}(E)$ and we
 assign this flux, calculated in $E$, to the energy $kE$. This means
 that: $\left(\frac{\ud N}{\ud E}(kE)\right)^\mathrm{new}=
 \frac{1}{k}\left(\frac{\ud N}{\ud E}(E)\right)^\mathrm{old}$ or
 $\left(k^3 E^3\frac{\ud N}{\ud E}(kE)\right)^\mathrm{new}=
 k^2\left(E^3\frac{\ud N}{\ud E}(E)\right)^\mathrm{old}$, where $k=1.15$
 for HiRes 1\&2 and $k=0.85$ for AGASA. These new, shifted, datasets
 agree quite well almost over the whole energy range as shown in 
 Fig.~\ref{fig:spectra190}.

\section{Results from the fits}

 We applied the method presented in the previous paragraph to the AGASA
 $-15\%$ data fitting the points in the energy range
 $10^{19}$--$10^{20}\eV$. The results are shown in the upper panel of
 Fig.~\ref{fig:spectra190} where we plot the best fits for $m=0,3,4$
 (solid lines). We only show results for $m$ = 3 and 4 because 
 these values bracket the cosmological evolution derived from 
 star forming regions and from gamma ray bursts. 
 The corresponding slopes are, respectively, $\alpha=2.55,2.45,2.45$
 with 1$\sigma$ errors of about 0.20. As it is clear from the plot
 all the three curves fit well the data in the region considered,
 with different degrees of goodness at low energy.
 The dotted lines represent the needed galactic component to fit the
 spectrum. The best fit with $m$=4 does not allow for galactic 
 component above $10^{17}\eV$.

 We repeated the same exercise for the HiRes $+15\%$ dataset 
 (which is shown in the lower panel of Fig.~\ref{fig:spectra190}) and
 the results were similar. The slopes are somewhat
 steeper, by about 0.05--0.1, well within the similarly
 large uncertainties.
 The main difference is at energies much lower than the fitting range,
 where the fits of the HiRes data set do not allow for 
 a galactic component in cases with cosmological evolution.
 The flux of extragalactic cosmic rays below $10^{19}\eV$ has
 to be slightly decreased by some additional process in order
 not to exceed the Akeno and HiRes measurements. 
 The shape of the spectra in the considered region
 is, however, the same for both experiments.

 It has to be noted that the inclusion of the error distribution 
 in the fit affects the spectral shape - the pile-up approaching
 $10^{20}\eV$ is visibly smoother. Since the points immediately
 above $10^{19}\eV$ have the lowest error bars, and thus affect
 the fit the most, the slope of the spectrum is increased by at most
 0.05, much smaller than the uncertainties from the fits.

 \begin{figure}
   \centering
   \includegraphics[width=85truemm]{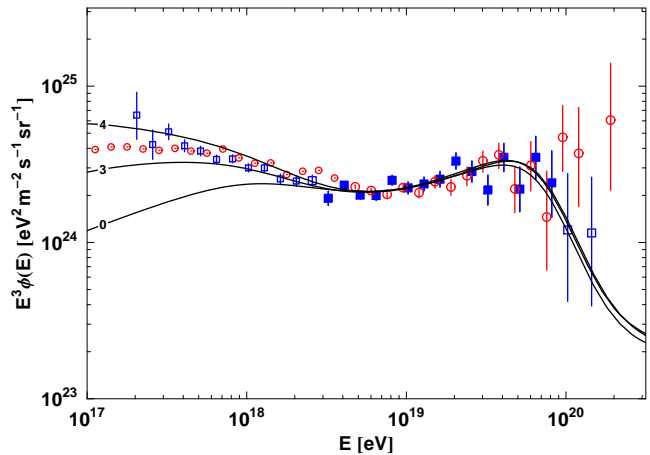}
   \vspace*{-10pt}
   \caption{Best fits to the HiRes $+15\%$ dataset in the
   $10^{18.5}$--$10^{20}\eV$ energy range. Symbols are the same 
   as in Fig.~\ref{fig:spectra190}.}
 \label{fig:hires185}
 \end{figure}

 The best fitting parameters depend slowly on the fitted energy range.
 Fig.~\ref{fig:hires185} shows the fit of the HiRes $+15\%$ data set
 for the energy range between $10^{18.5}$-$10^{20}\eV$ range. The
 best fit with $m$=4 again does not leave space for a galactic
 component below $10^{18}\eV$. The best spectral indices are 
 2.6, 2.5, and 2.5 respectively  for $m$ = 0, 3, and 4,
 slightly smaller than those for the higher fitting threshold and
 almost identical to the AGASA $-15\%$ set shown in 
 Fig.~\ref{fig:spectra190}. The 1$\sigma$ error bars decrease to
 about $\pm$0.1. The effect of the wider fitting range
 on the AGASA $-15\%$ spectrum is similarly small, although
 for this data set it increases the spectral slopes for all
 $m$ values by 0.05--0.1. The 1$\sigma$ fit errors decrease to slightly
 less than $\pm$0.1. 

  As a consistency check we also fitted
 the unmodified AGASA and HiRes spectra. The results we obtain are
 much like the ones presented above. The best fit parameters differ by
 about 0.05--0.1, with the same $\pm$0.2 errorbars.

\begin{figure}
   \centering
   \includegraphics[width=85truemm]{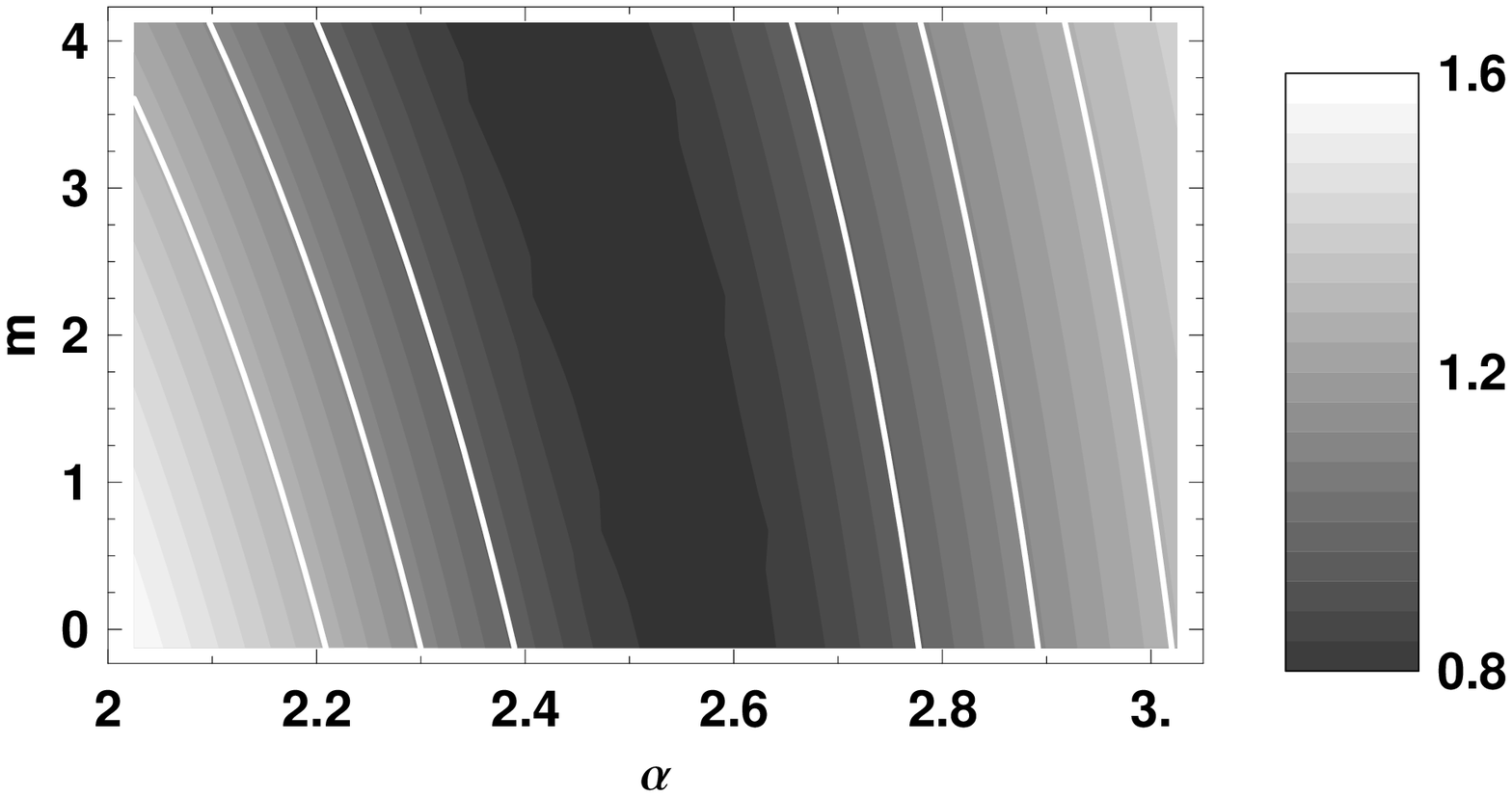}
   \includegraphics[width=85truemm]{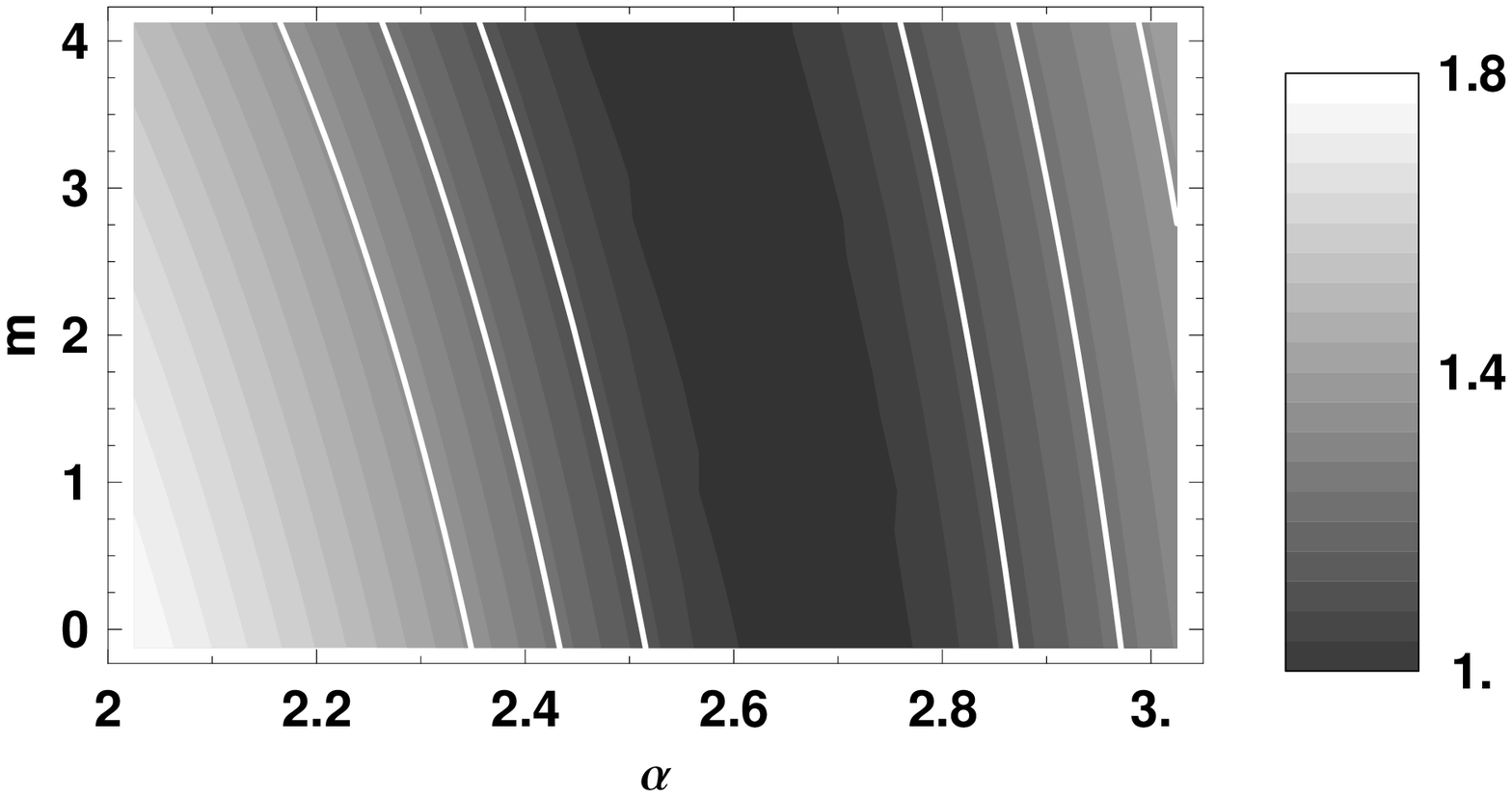}
  \vspace*{-10pt}
    \caption{Contour plots of $\log_{10}\chi^2$ as a function of $\alpha$
    and $m$ for the AGASA $-15\%$ dataset (upper panel) and for the 
    HiRes $+15\%$ data set (lower panel) fits above 10\protect$^{19}$ eV.
    The white lines are the contours
    corresponding to $1\sigma$, $2\sigma$ and
    $3\sigma$.}\label{fig:chi2}
  \vspace*{-10pt} 
 \end{figure}

 We performed several other fits varying the fitting threshold 
 between $10^{18.5}$ and $10^{19.2}\eV$ and convinced ourselves 
 that all fits returned consistent results within the 1$\sigma$
 errors of the presented fits as shown in Fig.~\ref{fig:chi2}.

 In this figure we plot in the top panel $\log_{10}\chi^2$ as a
 function of $(\alpha,m)$ for the AGASA $-15\%$ fit above $10^{19}\eV$,
 i.e. for six degrees of freedom. The analogous fitting of the
 HiRes $+15\%$ set is shown in the bottom panel.  
 The white contours are the
 confidence bands for $1\sigma$, $2\sigma$ and $3\sigma$.
 All $m$ values can provide a good fit to the data as the best fit
 $\alpha$ value slowly decreases with increasing $m$.
 This correlation is easily understood as with a
 smaller value of $\alpha$ less low energy particles
 are injected and to compensate for that one needs a
 stronger evolution of the sources to increase the
 number of low energy particles reaching the observer.
 For the HiRes $+15\%$ dataset the best fit parameters are in the strip
 connecting $(\alpha=2.6,m=0)$ and $(\alpha=2.4,m=4)$. As it is clear
 from the confidence bands in the plot, the present data sets do not
 restrict very much the values of the parameters, $\alpha$ being
 determined with an uncertainty of $\pm0.2$ for a given $m$ and $m$
 being almost free for a given $\alpha$.
 It is obvious, though, that
 fits with a flat injection spectrum do not give good $\chi^2$ values
 even with a strong cosmological evolution.
 Injection spectrum with $\alpha = 2.0$ would be in the 3$\sigma$ range
 only in the case of $m=4$. Flat injection
 spectrum models require that the galactic cosmic ray spectrum extends
 to $10^{19.5}\eV$. 

 If the shape of the cosmic ray  spectrum is the same as
 the one derived from the existing experimental  statistics, even much
 higher future statistics from the Auger observatory~\cite{Auger} would
 not help to solve it. We performed a fit with the current spectral
 shape and increased statistics that corresponds to the one expected
 from Auger. The 1$\sigma$ errors on $\alpha$ for the fits above
 $10^{19}\eV$ became only slightly narrower $\pm$0.15. The measurement
 of the cosmic ray chemical composition, or, a measurement of the flux
 of cosmogenic neutrinos generated by UHECR in propagation to
 us~\cite{ss05}, are needed to disentangle the two parameters. 

 \begin{figure}
   \centering
   \includegraphics[width=85truemm]{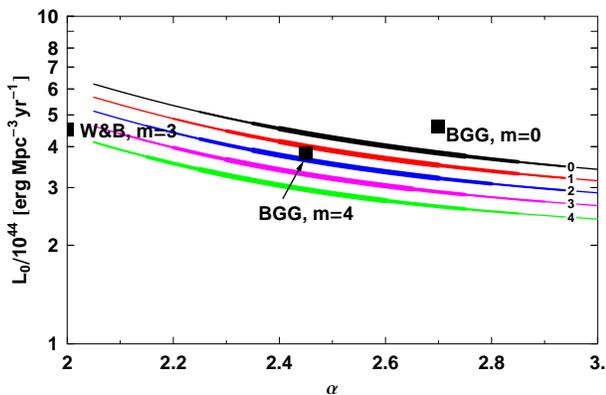}
  \vspace*{-10pt} 
  \caption{Required emissivity as a function of $(\alpha,m)$. Lines:
   best fit emissivity (above $10^{19}\eV$) as a function of $\alpha$.
   The numbers attached to the lines give the value of $m$. Every line is
   highlighted with different thicknesses corresponding to the confidence
   bands for $1\sigma$, $2\sigma$ and $3\sigma$. 
   The black squares are
   the best fits obtained in other works. W\&B corresponds to
   Ref.~\cite{WB,W95}, BGG corresponds to Ref.~\cite{BGG1}.}
 \label{fig:lum190}
  \vspace*{-15pt}
 \end{figure}
   
 In Fig.~\ref{fig:lum190} we plot the best fit present day emissivities
 above $10^{19}\eV$ as a function of $\alpha$ for different values of $m$.
 In this plot we also show the values obtained in
 Ref.~\cite{BGG1,WB,W95}.
 The differences with the results of Ref.~\cite{BGG1} are likely due to
 the slightly different dataset, to the different range of data used for
 the fits and to the inclusion in our calculation of the experimental
 energy resolution. There is also a factor because the fits were
 performed with a $-10\%$ shift of the AGASA data set instead of the
 $-15\%$ shift used here.
 It is interesting to note that the required values of the emissivity
 above $10^{19}\eV$ cover a narrow range between 2 and
 $6\cdot10^{44}\erg\Mpc^{-3}\yr^{-1}$ and that the required luminosity
 increases with the flattening of the injection spectrum. 
 This is a consequence of the fact that we only present the luminosity
 required above $10^{19}\eV$. If we were to extend the energy spectrum
 to lower energy, say to $10^{17}\eV$, we would observe exactly
 the opposite trend - steeper injection spectra would require much
 higher luminosity than flatter ones.

\section{Discussion and conclusions}
 
  After fitting the shifted AGASA and HiRes data sets in terms
 of injection spectral index and cosmological evolution of the 
 cosmic ray sources for an isotropic and homogeneous source
 distribution we obtained current emissivities above $10^{19}\eV$
 that differ only by about a factor of two. 
 In this sense we confirm the statement of Waxman~\cite{W95}
 that approximately the same emissivity is required for a 
 wide range on injection spectral indices. We disagree with
 the estimate of the central spectral index in Ref.~\cite{W95}
 and find a significantly steeper one. 
 
\begin{figure}
   \centering
   \includegraphics[width=85truemm]{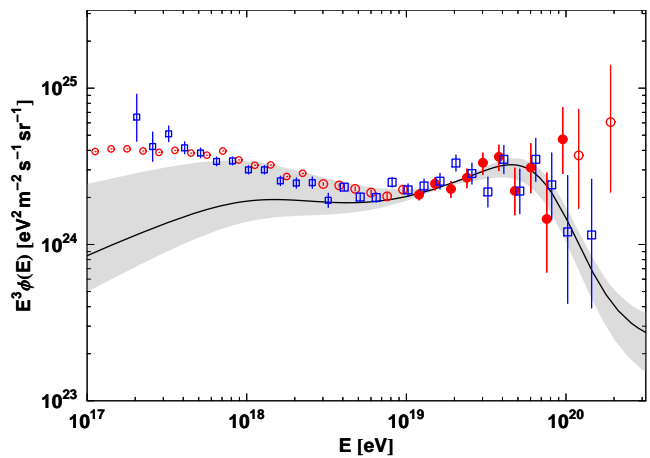}
 \vspace*{-10pt}
   \caption{ One \protect$\sigma$ error band (shaded)
   of the AGASA fit with \protect$m$=0 above 10\protect$^{19}$ eV.}
 \label{fig:agasa_m0}
 \vspace{-15pt}
\end{figure}
 
 Best fit spectral indices are, however, not well restricted
 by current statistics. In Fig.~\ref{fig:agasa_m0} we show
 with shaded area the 1$\sigma$ errors of the best fit 
 prediction from the AGASA data above $10^{19}\eV$
 (top panel of Fig.~\ref{fig:spectra190}) for $m$ = 0.
 The figure emphasizes the perils of all fits of the extragalactic 
 cosmic ray component with the current statistics. Such fits are
 most sensitive to, and attracted by, a small number of experimental
 points with the best statistics, in our case four points between 
 $10^{19}$ and $10^{19.4}\eV$. Fitting uncertainties do not affect
 much the higher energy spectra where the GZK suppression 
 prevails almost independently of the injection spectral index
 but create a large uncertainty below $10^{18.5}\eV$.
 This uncertainty makes estimates of the end of the galactic cosmic 
 ray spectrum by subtraction of the model predictions from the
 total observed flux unreliable.
 
  The luminosities that we show in Fig.~\ref{fig:lum190}
 apply only to energies above
 $10^{19}\eV$. If one is interested in the total cosmic ray
 luminosity of the cosmic ray sources one should continue
 the integration to much lower energies. This introduces 
 several possible new astrophysical parameters that come from
 the exact acceleration mechanism of the extragalactic cosmic rays.
 One could integrate down to the proton mass and obtain the 
 highest possible emissivity. On the other hand studies of
 particle acceleration at relativistic shocks~\cite{Achtetal} find
 a minimum acceleration energy of $m \Gamma_{\rm shock}^2$ which for
 $\Gamma_{\rm shock}$ = 1000, as in gamma ray bursts, could be $10^{15}\eV$
 and would decrease significantly the required emissivity.
 Modifications of the cosmic ray spectrum on propagation, such
 as suggested in Refs.~\cite{L04,AB04,parizot04} because of 
 magnetic horizon of the high energy cosmic rays~\cite{Stanev:2000fb}
 would not change the required emissivity.
 Such modifications would however change very much the
 extra-galactic cosmic ray spectrum suppressing the flux at the lower
 energy end and by consequence changing the shape of the end of the
 galactic cosmic ray spectrum required in order to fit the
 observations.
  
 Fits of the monocular HiRes data have been performed by the
 HiRes group~\cite{Bergman_fit}. The best fit is obtained for
 $\alpha$ = 2.38$\pm$0.04 and $m$ = 2.8$\pm$0.3 for a total of 42
 data points above $10^{17}\eV$ and respectively 39 degrees
 of freedom. 
  In additon to the different energy range of the fit, HiRes
  assumes a `toy' galactic cosmic ray model based on their
  composition measurement~\cite{HiRes_comp} that suggests
  domination of the extragalactic cosmic rays above 10$^{18}$ eV. 

 The HiRes fit is probably dominated by lower energy
 cosmic rays ($10^{17.5-18.5}\eV$) with much smaller error bars.
 The two fits give different central values for $\alpha$ and $m$ 
 but they are qualitatively consistent in the conclusion that even
 with a strong cosmological evolution of the cosmic ray sources 
 the observed spectra do not support flat injection spectra.

 We have fitted the shape of the ultrahigh energy cosmic ray
 spectrum above $10^{19}\eV$ assuming that these cosmic rays are
 protons, and that the sources of these protons are uniformly and
 homogeneously distributed in the Universe. The fits of the scaled AGASA
 and HiRes data sets allow for power law injection spectra in the range
 $A E^{-(2.4-2.7)}$ for cosmological evolution of the cosmic ray sources
 between $(1+z)^4$ to $(1+z)^0$.
 The cosmic ray emissivities above $10^{19}\eV$ required by
 different models are within about a factor of two in this range.
 The best fit spectral index decreases
 for strong evolution models. Flatter injection spectra do not
 fit well the cosmic ray spectra above $10^{19}\eV$. This also
 means that the end of the galactic cosmic ray spectrum is at,
 or below, $10^{18}\eV$ depending on the cosmological evolution
 of the extragalactic cosmic ray sources. 
 Consistent data on the cosmic ray composition in the energy
 range above $10^{17}\eV$ are required in order to reveal the
 end of the galactic cosmic ray spectrum and thus help determine
 that of the extragalactic sources.

{\bf Acknowledgments}. We thank D.~Seckel for useful discussions.
This research is funded in part by NASA APT grant NNG04GK86G.

\end{document}